\documentclass[preprint,showpacs,preprintnumbers,amsmath,amssymb]{revtex4}
\usepackage{graphics}
\usepackage{epsfig}
\usepackage{dcolumn}% Align table columns on decimal point
\usepackage{bm}% bold math
\begin{document}

%\preprint{IFT-P.09/2007}
%\preprint{ArXiv:1002.xxxx}
\title{Asymmetries in $e^+e^-\to f\bar{f}$ processes at  ILC\\ for
models with an extra neutral vector boson $Z^\prime$}

% \altaffiliation[Also at ]{Physics Department, XYZ
%University.}%
\author{E. C. F. S. Fortes}%
\email{elaine@ift.unesp.br}
\affiliation{ Instituto  de F\'\i
sica Te\'orica--Universidade Estadual Paulista \\ R. Dr. Bento
Teobaldo Ferraz 271, Barra Funda\\ S\~ao Paulo - SP, 01140-070,
Brazil}
\author{J. C. Montero}%
\email{montero@ift.unesp.br}
\affiliation{ Instituto  de F\'\i
sica Te\'orica--Universidade Estadual Paulista \\ R. Dr. Bento
Teobaldo Ferraz 271, Barra Funda\\ S\~ao Paulo - SP, 01140-070,
Brazil}
\author{V. Pleitez}%
\email{vicente@ift.unesp.br}
\affiliation{ Instituto  de F\'\i
sica Te\'orica--Universidade Estadual Paulista \\ R. Dr. Bento
Teobaldo Ferraz 271, Barra Funda\\ S\~ao Paulo - SP, 01140-070,
Brazil}

\date{25/11/2010}% It is always \today, today,

\begin{abstract}
Many extensions of the standard model predict the
existence of extra neutral vector bosons, generically referred as $Z^\prime$.
This boson may be discovered at the LHC but in this case it will be necessary to study the
respective parameters in order to discriminate to which model it belongs to. This is a task for a much clean
lepton linear collider as the future ILC. In this paper we develop an exemplary
study of the capability of several asymmetries on and off $Z^\prime$ peak in
discriminating among those extensions with (almost) no ambiguities.

\begin{center}
Version to be published in Physical Review D
\end{center}

%In this paper we study the capability of several asymmetries, on- and off- the $Z^\prime$ peak,
%that can be measured in that sort of collider  in discriminating among those extensions with (almost) no ambiguities.

\end{abstract}

\pacs{13.66.Hk, 14.70.Pw, 12.60.Cn
}

\maketitle

\section{Introduction}
\label{sec:intro}

Among the most motivated extensions of the standard model (SM) are those in which the gauge
symmetries of the model are extended by at least one extra $U(1)$ factor, probably, a remnant
of unknown physics at higher energy scales. In fact, these sort of models arise, for example,
in grand unified theories inspired, or not, in superstring, left-right models, and supersymmetric
models~\cite{langacker,susyu1}. An interesting feature of these models is
the existence of extra neutral vector bosons, generically denoted by $Z^\prime$, which may be
manifested at the TeV scale, implying a rich phenomenology in the hadron and future lepton
colliders~\cite{zprime}. These bosons appear also in models of dynamical symmetry breaking,
Little Higgs models, models with extra dimensions, or even through the Stueckelberg mechanism which
avoids the Higgs mechanism for generating the mass for the $Z^\prime$ boson~\cite{yuca}. If they do exist
they can be discovered at the LHC but this will only be the first step: next it will be necessary
to discriminate the model that better fits all the measured parameters as its mass,
its partial and total widths, and its couplings to the known fermions, etc. The detailed study of
these properties can be done at a linear collider like the ILC.

Concerning the quantum numbers related to the Abelian factors, the options are numerous but some most
motivated ones involve the baryon ($B$), total lepton ($L$), or lepton
family ($L_i$), numbers, or combinations of some of them. However, it is still possible that
they are related with an exotic charge ($X$). In particular, it is interesting that the extra local
$U(1)$ symmetry is related to $B-L$ since $U(1)_{_{B-L}}$ is an automatic global  symmetry  of the
degrees of freedom of the SM i.~e., without right-handed neutrinos and it is also anomalous~\cite{thooft}.
It is possible to gauge this symmetry when an appropriate number of right-handed neutrinos is introduced,
in order to make this symmetry anomaly free. Moreover, the energy scale in which
the new $U(1)_{B-L}$ symmetry is broken is related to the generation of the light neutrino masses through
the seesaw mechanism. The simplest way to implement $B-L$ as a gauge symmetry is by adding just an extra $U(1)$
factor, which generator may, or may not, commute with the $U(1)_Y$ generators. In the first case,
the electric charge operator has also a component on this factor, and in the second
one, the electric charge operator is the same as in the SM, i.~e., it
has no component on the extra $U(1)$ symmetry. We call the former sort of models
``flipped", as the models in Ref.~\cite{blsm}, and ``secluded" models the latter ones~\cite{erler0},
and an example (nonsupersymmetric) is in Ref.~\cite{appelquist}. Although usually this
extra symmetry is related to a grand unified theory, from the
phenomenological point of view, we can build models in this simplest way,
independently of their origin, and study their properties in lepton or hadron
colliders.

In general, the masses of the new neutral vector boson must be in the order of few TeV, or be very
weakly coupled to the known matter, in order to maintain
consistency with the present phenomenology. This boson may be discovered at the LHC, but in order to
study its respective parameters a linear leptonic collider, like the proposed ILC~\cite{ilc,lhcilc},
is better suited. Below, we will show that the study of several asymmetries, on- and off-
the $Z^\prime$ peak, in this type of collider can be used to discriminate among the different models
which have this extra neutral vector boson.
 Among the most studied models in the literature are those
based on the $E_6$ GUT group and left-right symmetry groups~\cite{erler}, and also those in which
an extra $U(1)_{_{B-L}}$ factor is added, as those discussed above.

The outline of this paper is as follows. In Sec.~\ref{sec:models} we give a
general discussion of the models whose details are presented in Secs.~\ref{subsec:flipped}
and \ref{subsec:secluded}, $E_6$ and left-right symmetric models are summarized in Sec.~\ref{subsec:e6lr}.
Our results and discussions appear in Sec.~\ref{sec:results},
and the conclusions in Sec.~\ref{sec:con}. Exact analytical expressions for the neutral
current coefficients defined in Eq.~(\ref{nc}), for the flipped  model
are given in the Appendix \ref{sec:neutral1}, while in Appendix~\ref{sec:widths} we show
the partial decay widths of the neutral vector boson mass eigenstates, denoted by $Z_{1,2}$.

\section{The models}
\label{sec:models}

Here we will consider extensions of the
electroweak standard model in which there are two $U(1)$ gauge factors, i.~e.,
these models are based on the $SU(2)_L\otimes U(1)_1\otimes U(1)_2$ gauge symmetries.
In general, there is a mixing between the two Abelian
gauge bosons in the kinetic term and in the mass term~\cite{delaguila2};
however, we will work in a basis in which the kinetic term, at the tree level, is
already diagonal,  and the mixing between the two neutral vector bosons
may appear, or not, in the mass term.

Below we will denote $Z$ and $Z^\prime$ the symmetry
eigenstates, and $Z_1$ and $Z_2$ the mass eigenstates. However, if the mixing
between $Z$ and $Z^\prime$ is small, then $Z\approx Z_1$ and $Z^\prime\approx Z_2$.
It is possible, as in the secluded pure $B-L$ model that $Z_1\equiv Z$ and $Z_2\equiv Z^\prime$.
We will parameterize the neutral currents in terms of the mass eigenstate
fields as follows~\cite{pdg}:
\begin{equation}
{\mathcal{ L}}^{NC}=- \frac{g}{2c_W}\sum_i  \overline{ \psi_i}
\gamma_\mu [(g^i_V-g^i_A \gamma_5 ) Z^\mu_1 + (f^i_V-
f^i_A\gamma_5 )Z^\mu_2]\psi_i.
\label{nc}
\end{equation}

The couplings of $Z_1$ with fermions are defined as $g_V=(1/2)(g_L+g_R)$ and $g_A=(1/2)(g_L-g_R)$, where $g_L$
and $g_R$ are the dimensionless coupling constants of the left-(right-) handed
fermions. Similar definitions exist for the $f_V,f_A$ and $f_L,f_R$ couplings related
to $Z_2$.

In this paper we present a detailed study of several asymmetries
in models which have an extra neutral vector boson $Z^\prime$ in the context of a linear leptonic collider. We show that
the measurement of these asymmetries will allow the determination of the parameters
related to the extra neutral vector bosons, if they are discovered at LHC, and the
possibility to distinguish the models from each other.

\subsection{The flipped $U(1)_{_{B-L}}$ model}
\label{subsec:flipped}

The first model is based on the following electroweak gauge symmetry~\cite{blsm}:
\begin{eqnarray}
SU(2)_L\otimes U(1)_{Y^\prime}\otimes U(1)_{B-L}\to
SU(2)_L\otimes U(1)_Y \to U(1)_{em},
\label{group1}
\end{eqnarray}
where $Y^\prime$ is chosen to obtain the hypercharge $Y$ of the
standard model, given by $Y=~Y^\prime+~(B-L)$. Thus, in this
case, the electric charge operator $Q$ is given by
\begin{equation}
    \frac{Q}{e}=I_3+\frac{1}{2}\,\left[Y^\prime +
    (B-L)\right].
    \label{gn1}
\end{equation}
There are several versions of model depending on the lepton
number attributed to the right-handed neutrinos, see
Ref.~\cite{blsm} for details.
Here we will consider a model in which we add one right-handed neutrino per generation,
$n_{\alpha R},\,\alpha=e,\mu,\tau$, and a complex scalar singlet, $\phi$, to the usual
representation content of the SM. The quantum numbers of the
degrees of freedom for the model are given in Table I of Ref.~\cite{blsm}. The
baryon number assignment is as usual. The scalar sector consists of a Higgs doublet
$\Phi=(\varphi^+\,\varphi^0)^T$ and the singlet $\phi$.
The neutral gauge bosons $W^\mu_3$, $B^\mu_{Y^\prime}$, and $B^\mu_{_{B-L}}$,
corresponding to the $SU(2)_L$, $U(1)_{Y^\prime}$, and $U(1)_{B-L}$ gauge factors
respectively, are mixtures of the photon, $A^\mu$, and two
massive neutral bosons, $Z^\mu_1$, and $Z^\mu_2$, fields.

Denoting $\langle\varphi^0\rangle=v/\sqrt{2}$, the Higgs doublet vacuum
expectation value (VEV) $v\sim246$ GeV, and
$\langle\phi\rangle=u/\sqrt{2}$, the VEV of the neutral
scalar singlet, the interesting parts of the covariant derivative are
\begin{equation}
\frac{v^2}{8}(-gW^\mu_3+g^\prime
B^\mu_{Y^\prime})^2+\frac{u^2}{8}
(g^{\prime}Y^\prime_\phi B^\mu_{Y^\prime}-
g_{_{B-L}}Y^\prime_\phi B^\mu_{_{B-L}})^2,
\label{derivada1}
\end{equation}
and we will choose $Y^\prime_\phi=-2$ since for the scalar singlet $Y^\prime_\phi=-(B-L)$.
Notice that in this case the right-handed neutrinos may obtain a
renormalizable Majorana mass term; if we had chosen
$Y^\prime_\phi=-1$ only dimension five Majorana mass terms would be
allowed and for $Y^\prime_\phi>2$ the singlet $\phi$ couples to right-handed neutrinos
only through higher dimension operators. The square mass matrix for the neutral vector bosons
in the $W_3,B_{Y^\prime}, B_{_{B-L}}$ basis is given by
\begin{equation}
M^2_{\textrm{neutral}}=g^2\,u^2\,\left(
                  \begin{array}{ccc}
                    \bar{v}^2/4 & -t^\prime \bar{v}^2/4& 0 \\
                    -t^\prime \bar{v}^2/4 &
                    t^{\prime\,2}(1+\bar{v}^2/4) &
                    -t^\prime t_{_{B-L}} \\
                    0 & -t^\prime t_{_{B-L}}& t^2_{_{B-L}} \\
                  \end{array}
                \right),
                \label{msquare1}
\end{equation}
where we have defined $t^\prime=g^\prime/g$,
$t_{_{B-L}}=g_{_{B-L}}/g$ and $\bar{v}=v/u$; and
$\textrm{Det}\,M^2_{\textrm{neutral}}~=~0$ as must be.

The exact mass eigenvalues are: zero for the photon field, and
\begin{equation}
M^2_{1,2}~=~\frac{g^2u^ 2}{8}\left(A\mp \sqrt{ A^
2-16B\bar{v}^2}\right),
\label{massvector1}
\end{equation}
for the two massive vector fields; where we have defined
\begin{eqnarray}
 A=4(t^{\prime\,2}+t^
 2_{_{B-L}})+(1+t^{\prime\,2})\bar{v}^2,\quad
 B=t^{\prime\,2}(1+t^2_{_{B-L}})+t^2_{_{B-L}}.
\label{ab}
\end{eqnarray}

In the approximation $\bar{v}\ll1$ the mass eigenvalues are
given by
\begin{eqnarray}
M^2_1&\approx & g^2\,\frac{v^2}{4}\left(1+ \frac{t^{\prime\,2}
t^2_{_{B-L}}} {t^{\prime\,2}+t^2_{_{B-L}}}
-\frac{t^{\prime\,4}}{4c^2_W}\,\bar{v}^2\right) \nonumber \\
&=& \frac{g^2v^2}{4c^2_W}\,\left(1-\frac{t^{\prime\,4}\bar{v}^2
}{4}\, \right),\nonumber \\ M^2_2 &\approx
&g^2u^2(t^{\prime\,2}+t^2_{_{B-L}})\,\left(1 + \frac{
t^{\prime\,4}}{ 4 (t^{\prime\,2}+t^2_{_{B-L}})^2}
\,\bar{v}^2\right).
\label{massaapprox1}
\end{eqnarray}
where we have used Eq.~(\ref{thetaw}) below in the first line
of Eq.~(\ref{massaapprox1}) showing explicitly that at the
order $v^2/u^2$ there is not dependence on $g_{_{B-L}}$ in the
lightest neutral vector boson.

The relation among the $U(1)$ charges is
\begin{equation}
\frac{g^2}{e^2}=1+\frac{1}{t^{\prime\, 2} }+\frac{1}{
t^2_{_{B-L}} },
\label{aquela}
\end{equation}
and the electroweak mixing angle is given by
\begin{equation}
t^2_W=\frac{t^{\prime\,2} t^2_{_{B-L}}  }{
t^{\prime\,2} +t^2_{_{B-L}} },\;\;s^2_W=
\frac{t^{\prime\,2}t^2_{_{B-L}}}{t^{\prime\,2}(1+t^2_{_{B-L}})
+t^2_{_{B-L}}},%\;\;c^2_W=\frac{t^{\prime\,2}
%+t^2_{_{B-L}}}{t^{\prime\,2}(1+t^2_{_{B-L}})+t^2_{_{B-L}} },
\label{thetaw}
\end{equation}
where $t^2_W\equiv \tan^2\theta_W\equiv g^2_Y/g^2$, etc. Notice that since
$g^2/g^2_Y=1/t^{\prime \,2}+1/t^2_{_{B-L}}$, where $g_Y$ is
the standard model $U(1)_Y$ coupling constant, it means that
\begin{equation}
t^{\prime\,2}=\frac{t^2_W}{1-\frac{t^2_W}{t^2_{_{B-L}} }},
\quad t^2_{_{B-L}}=\frac{t^2_W}{ 1-\frac{t^2_W}{t^{\prime 2}} },
\label{constraint}
\end{equation}
which implies that $g^\prime,g_{_{B-L}}>g_Y=e/\cos\theta_W=g\tan\theta_W$.

For the eigenvectors, we find that the one corresponding to the zero mass
eigenvalue, the photon, is independent of the VEV
structure, and is given exactly by:
\begin{equation}
\mathcal{A} = \frac{1}{\left(1+\frac{1}{t^{\prime\,2}}+\frac{1}{
t^2_{_{B-L}} } \right)^{ 1/2}
}\left(1,\,\frac{1}{t^\prime},\,\frac{1}{t_{_{B-L}} }
  \right),
\label{foton1}
\end{equation}
while for the massive ones, that we will write explicitly only in the case when $u\gg
v$, the normalized eigenvectors are:
\begin{eqnarray}
Z_1 &\approx&
  \frac{c_W}{(t^{\prime\,2}+t^2_{_{B-L}})}\,  \left(
  -(t^{\prime\,2}+t^2_{_{B-L}}),\,t^\prime
  t^2_{_{B-L}},\,t^{\prime\,2}
  t_{_{B-L}}\right),\\
  Z_2  &\approx &
  \frac{1}{(t^{\prime\,2}+t^2_{_{B-L}})^{1/2}}
  \left(0\,,-t^\prime\,,t_{_{B-L}}\right).
  \label{z1z21}
\end{eqnarray}
Only in this approximation these eigenvectors are independent of the VEVs.

The neutral current couplings of $Z_{1,2}$ with the known fermions are
parameterized as in Eq.~(\ref{nc}). The exact coefficients, $g^i_{V,A}$ and
$f^i_{V,A}$, that are given in Appendix~\ref{sec:neutral1}, were calculated by
using the full analytical expressions for $Z_{1,2}$. Below, we shown them in the
approximation $\bar{v}^2\ll 1$.

The couplings of the neutrinos are
\begin{eqnarray}
&&g^\nu_V \approx
\frac{1}{2}+\frac{t^{\prime\,2}(t^{\prime\,2}+2t^2_{_{B-L}})}{8(t^{\prime\,2}
+t^2_{_{B-L}})^2}\,\bar{v}^2,
\quad\quad g^\nu_A \approx
\frac{1}{2}-\frac{t^{\prime\,4}}{8\,(t^{\prime\,2}+t^2_{_{B-L}})^2}\,\bar{v}^2,\nonumber
\\&& f^\nu_V \approx
\frac{t^{\prime\,2}+2t^2_{_{B-L}}}{2t^{\prime}t_{_{B-L}}
}\,s_W-
\frac{t^{\prime\,3}t_{_{B-L}}}{8(t^{\prime\,2}+t^2_{_{B-L}})^2
}\,\frac{1}{s_W}\,\bar{v}^2, \nonumber \\ &&f^\nu_A\approx
-\frac{t^\prime}{2t_{_{B-L}}}\,s_W-\frac{t^{\prime\,3}t_{_{B-L}}}{8(g^{\prime\,2}
+t^2_{_{B-L}})^2}\,
\frac{1}{s_W}\,\bar{v}^2.
\label{nusnosso}
\end{eqnarray}

For the case of the charged leptons we have
\begin{eqnarray}
&&g^l_V\approx-\frac{1}{2}+2s^2_W-\frac{t^{\prime\,2}(t^{\prime\,2}
-2t^2_{_{B-L}})
}{8(t^{\prime\,2}+t^2_{_{B-L}})^2 } \;\bar{v}^2,\quad\quad
g^l_A\approx -g^\nu_A
%\frac{1}{2}+\frac{g^{\prime\,4}}{8(g^{\prime\,2}+g^2_{_{B-L}})^2}\,\bar{v}^2.
\nonumber \\&& f^l_V\approx
-\frac{1}{2}\,\frac{t^{\prime\,2}-2t^2_{_{B-L}}} {t^\prime
t_{_{B-L}}}\,s_W + \frac{t^{\prime\,3}t_{_{B-L}} }
{8(t^{\prime\,2}+t^2_{_{B-L}})^2 } \left(
\frac{1-4s^2_W}{s_W}\right)\,\bar{v}^2, \nonumber \\&&
f^l_A\approx \frac{t^\prime}{2t_{_{B-L}}}\,\left(1+
 \frac{t^{\prime\,2}t^2_{_{B-L}} }
 {4(t^{\prime\,2}+t^2_{_{B-L}})^2\,s^2_W}\,\bar{v}^2\right)s_W.
\label{lnosso}
\end{eqnarray}

In the quark sector we obtain, for the $u$-like quarks
\begin{eqnarray}
&&g^u_V\approx \frac{1}{2}-\frac{4}{3}s^2_W+
\frac{t^{\prime\,2}(3t^{\prime\,2}-2t^2_{_{B-L}})}
{24(t^{\prime\,2}+t^2_{_{B-L}})^2}\,\bar{v}^2,\quad\quad
g^u_A\approx  -g^l_A, \nonumber \\&& f^u_V\approx\frac{
3t^{\prime\,2}-2t^2_{_{B-L}} } {6t^\prime t_{_{B-L}}
}\,s_W+\frac{t^\prime } {24t_{_{B-L}}}
\left(5t^{\prime\,2}t^2_{_{B-L}}-3t^2(t^{\prime\,2}+t^2_{_{B-L}})\right)\,s_W\,\bar{v}^2,
\nonumber \\&& f^u_A\approx
-\frac{t^\prime}{2t_{_{B-L}}}\,\left(1+
 \frac{t^{\prime\,2}t^2_{_{B-L}} }
 {4(t^{\prime\,2}+t^2_{_{B-L}})^2\,s^2_W}\,\bar{v}^2\right)s_W,
\label{unosso}
\end{eqnarray}
and, for the $d$-like quarks
\begin{eqnarray}
&&g^d_V\approx-\frac{1}{2}+\frac{2}{3}s^2_W-\frac{t^{\prime\,2}(3t^{\prime\,2}
+2t^2_{_{B-L}})} {24(t^{\prime\,2}+t^2_{_{B-L}})^2}\,\bar{v}^2,
\quad\quad g^d_A\approx g^l_A, \nonumber \\
&&f^d_V\approx-\frac{3t^{\prime\,2}+2t^2_{_{B-L}} } {6t^\prime
t_{_{B-L}}}\,s_W +\frac{t^\prime}{24
(t^{\prime\,2}+t^2_{_{B-L}})^2t_{_{B-L}} } \left(
-t^{\prime\,2}t^2_{_{B-L}}+3t^2(t^{\prime\,2}+t^2_{_{B-L}})\right)\,s_W\,\bar{v}^2,
\nonumber \\&& f^d_A\approx -f^u_A.
\label{dnosso}
\end{eqnarray}

We see from Eqs.~(\ref{nusnosso})-(\ref{dnosso}) that the
$g_{V,A}$ couplings are that of the SM, at the tree level, plus
corrections that are suppressed by the $\bar{v}^2$ factor, i.~e.,  for $t^\prime$
and $t_{_{B-L}}$ fixed if $\bar{v}\to 0$, $g_{V,A}$ go to the tree level standard model
expressions and the couplings $f_{V,A}$ depend only on
$s_W,g^\prime$, and $g_{_{B-L}}$.

\subsection{The secluded $U(1)_z$ model}
\label{subsec:secluded}

The other electroweak model is  based on the gauge symmetry
\begin{eqnarray}
SU(2)_L\otimes U(1)_Y\otimes U(1)_z \to SU(2)_L\otimes U(1)_Y
\to U(1)_{em},
\label{group2}
\end{eqnarray}
where $Y$ is the weak hypercharge, and $Q$ is given as usual,
\begin{equation}
    \frac{Q}{e}=I_3+\frac{1}{2}\,Y.
    \label{gn2}
\end{equation}
In this case the electric charge has no component in $U(1)_z$. Depending on the
$U(1)_z$ charge of the Higgs scalar we can have several versions of this
model~\cite{appelquist}.
There exist other solutions to the anomaly cancellation equations
if a second Higgs doublet is introduced~\cite{carena}.

The scalar doublet $H$ carries $Y=+1$, as usual. In addition, the model has a neutral complex scalar singlet
$\varphi$ carrying only the $U(1)_z$ charge which is equal to $+2$.  We use again
$\langle H^0\rangle=v/\sqrt{2}$ and $\langle\varphi\rangle=u/\sqrt{2}$.

%\subsection{The vector boson sector}
%\label{subsec:vector2}

In this model the mass square matrix for the neutral gauge
bosons arises from the terms
\begin{equation}
g^2\frac{v^2}{8}(W^\mu_3-t_W B^\mu_Y-z_Ht_zB^\mu_z)^2+
g^2\frac{u^2}{8}(z_\varphi t_zB^\mu_z)^2, \label{derivada3}
\end{equation}
in the covariant derivatives; we have defined $t_W=g_Y/g$ and $t_z=g_z/g$.
Here we will assume $z_\varphi=2$ and not $z_\varphi=1$ as in Ref.~\cite{appelquist}.
We will consider $z_H=0$ only, since in this case $U(1)_z\equiv U(1)_{_{B-L}}$.
In this case, the mass square of the two neutral vector bosons are
\begin{equation}
M^2_1=\frac{1}{4}g ^2(1+t^2_W)v^2\equiv M^2_Z,\quad
 M^2_2=g^2t^2_zu^2=g^2_zu^2,
\label{massvector4}
\end{equation}
and $g^i_{V,A}$ are exactly the same as the
standard model at the tree level, $f^i_A=0$ and the $f_V$s are given by
\begin{equation}
f^\nu_V=f^l_V=-3f^u_V=-3f^d_V=t_z\,c_W=g_z\,\frac{c_W}{g}.
\label{fvfazh0}
\end{equation}
Notice that the $Z_2$ couples universally with all
fermions with strength $g_z$ since the factor $c_W/g$ appears only
due to the parametrization in Eq.~(\ref{nc}). Notice that when $z_H=0$,
$t_z$  is in fact $t_{_{B-L}}$.

\subsection{$E_6$ and left-right symmetric models}
\label{subsec:e6lr}

The other models that we will consider here are those in which the electroweak effective
gauge symmetries are $SU(2)_L\otimes U(1)_Y\otimes U(1)_X$  and also the left-right symmetric model.
In the former models, the charge $X$ is related to a grand unified theory in which the model is embedded,
for instance, in $E_6$ models if the following breaking chain occurs:
\begin{eqnarray}
E_6&\to& SO(10)\otimes U(1)_\chi\nonumber \\&\to& SU(5)\otimes U(1)_\psi\otimes U(1)_\chi\nonumber \\ &\to&
SU(3)\otimes SU(2)_L\otimes U(1)_Y\otimes U(1)_X,
\label{group3}
\end{eqnarray}
with $U(1)_X=U(1)_\chi\cos\beta+U(1)_\psi\sin\beta$,
when $\beta=0,\pi/2,-\arctan(\sqrt{5/3})$ we have, respectively, the pure $U(1)_\chi$, $U(1)_\psi$
and $U(1)_\eta$ model.
Next, we consider the effective model based on $SU(2)_L\otimes U(1)_Y\otimes U(1)_X$ originated
from the breaking of the $SU(2)_L\otimes SU(2)_R\otimes U(1)_{_{B-L}}$.
left-right symmetric model, $g_L=g_R$. The respective coefficients $f_V$ and $f_A$
are shown in Table~\ref{table1}~\cite{dittmar}.

\begin{table}[ht]
\begin{center}
\begin{tabular}{|l||c|c||c|c|}
\hline
& \multicolumn{2}{|c||}{$E_6$} & \multicolumn{2}{|c||}{LR} \\ \cline{2-5}
%\hline
fermion& $f_V$ & $f_A$ &  $f_V$ & $f_A$  \\
\hline
neutrinos & $\frac{3c_\beta}{4\sqrt{6}}+\frac{\sqrt{10}s_\beta}{24}$ & $=f_V$
&$\frac{1}{4\alpha_{LR}}$ & $=f_V$ \\ \hline
leptons & $\frac{c_\beta}{\sqrt{6}}$ & $\frac{c_\beta}{2\sqrt{6}}+\frac{\sqrt{10}s_\beta}{12}$
& $\frac{1}{2\alpha_{LR}}-\frac{\alpha_{LR}}{4}$	  &	$\frac{\alpha_{LR}}{4}$      \\ \hline
$u$-quarks & 0 & $-\frac{c_\beta}{2\sqrt{6}}+\frac{\sqrt{10}s_\beta}{12}$  &
$-\frac{1}{6\alpha_{LR}}+\frac{\alpha_{LR}}{4}$   &    $-\frac{\alpha_{LR}}{4}$     \\ \hline
$d$-quarks & $-\frac{c_\beta}{\sqrt6}$  & $\frac{c_\beta}{2\sqrt6}+\frac{\sqrt{10}s_\beta}{12}$
&$-\frac{1}{6\alpha_{LR}}-\frac{\alpha_{LR}}{4}$  &$\frac{\alpha_{LR}}{4}$     \\
\hline
\end{tabular}
\end{center}
\vskip -0.5cm
\caption{
Vector and axial-vector couplings of the $Z_2$ in $E_6$ inspired models. The values $\beta=0,\pi/2,\arctan(-\sqrt{5/3})$,
correspond to $Z^\prime_\chi,Z^\prime_\psi$ and $Z^\prime_\eta$ models, respectively.
It is also shown the respective couplings in the left-right model.
}
\label{table1}
\end{table}

%%%
\begin{figure}[ht]
\begin{center}
\includegraphics[height=.5\textheight]{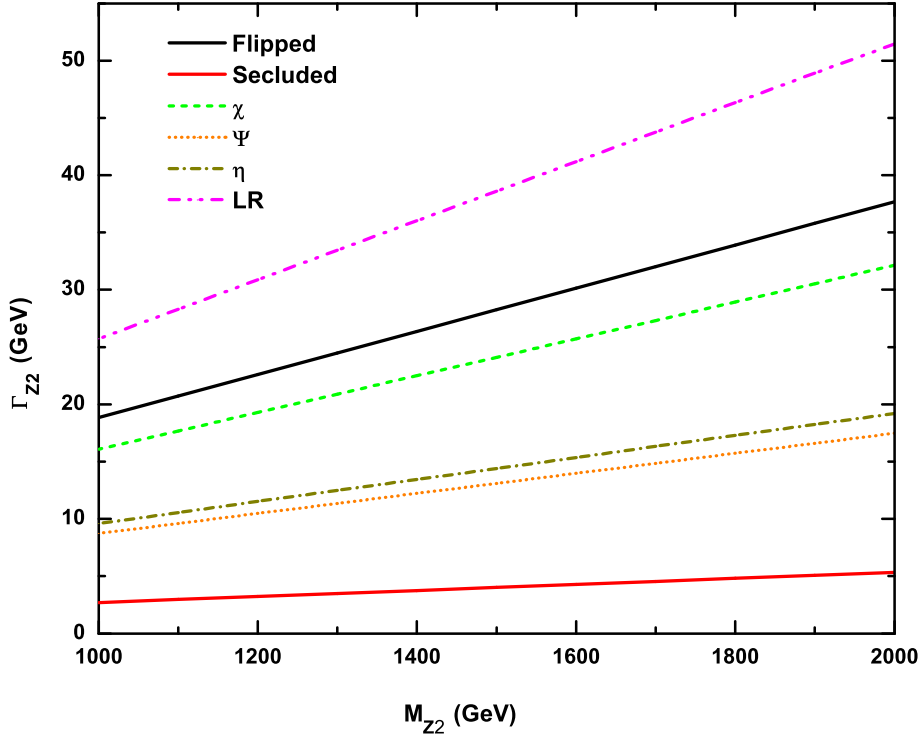}
\caption{\label{fig1}Evolution of the $Z_2$ decay width with its mass. }
\end{center}
\end{figure}

%\figura{12.38}{9.6}{Larguratodos.eps}{\label{fig1} Evolution of the $Z_2$ decay
%width with its mass.
%}

%The discovery potential of the $Z_2$ and the production of heavy neutrinos at
%LHC, for the secluded $B-L$ ($z_H=0$) model, were considered
%in Ref.~\cite{basso}.

\begin{figure}[ht]
\begin{center}
\includegraphics[height=.5\textheight]{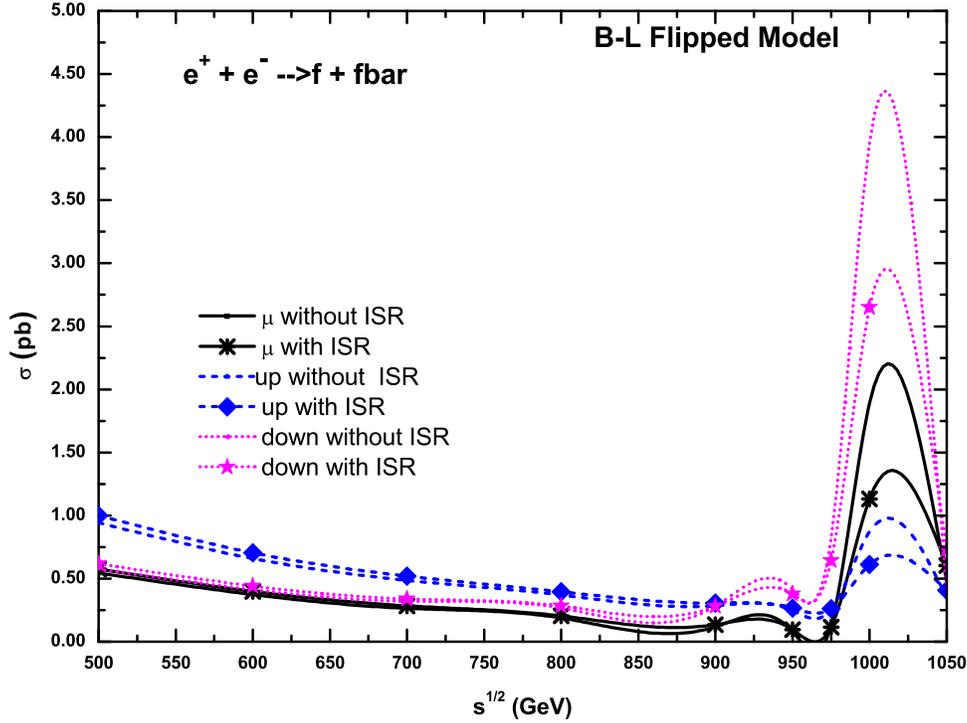}
\caption{\label{figa} Cross sections
for the $e^+e^-\to f\bar{f}$ process in the flipped model with and without
beamstrahlung (initial state radiation) corrections. }
\end{center}
\end{figure}

%\figura{12.38}{9.6}{crossallflip.eps}{\label{figa} Cross sections
%for the $e^+e^-\to f\bar{f}$ process in the flipped model with and without
%beamstrahlung (initial state radiation) corrections.%{\color{red}Ver observa\c{c}\~{a}o }
%}

\begin{figure}[ht]
\begin{center}
\includegraphics[height=.5\textheight]{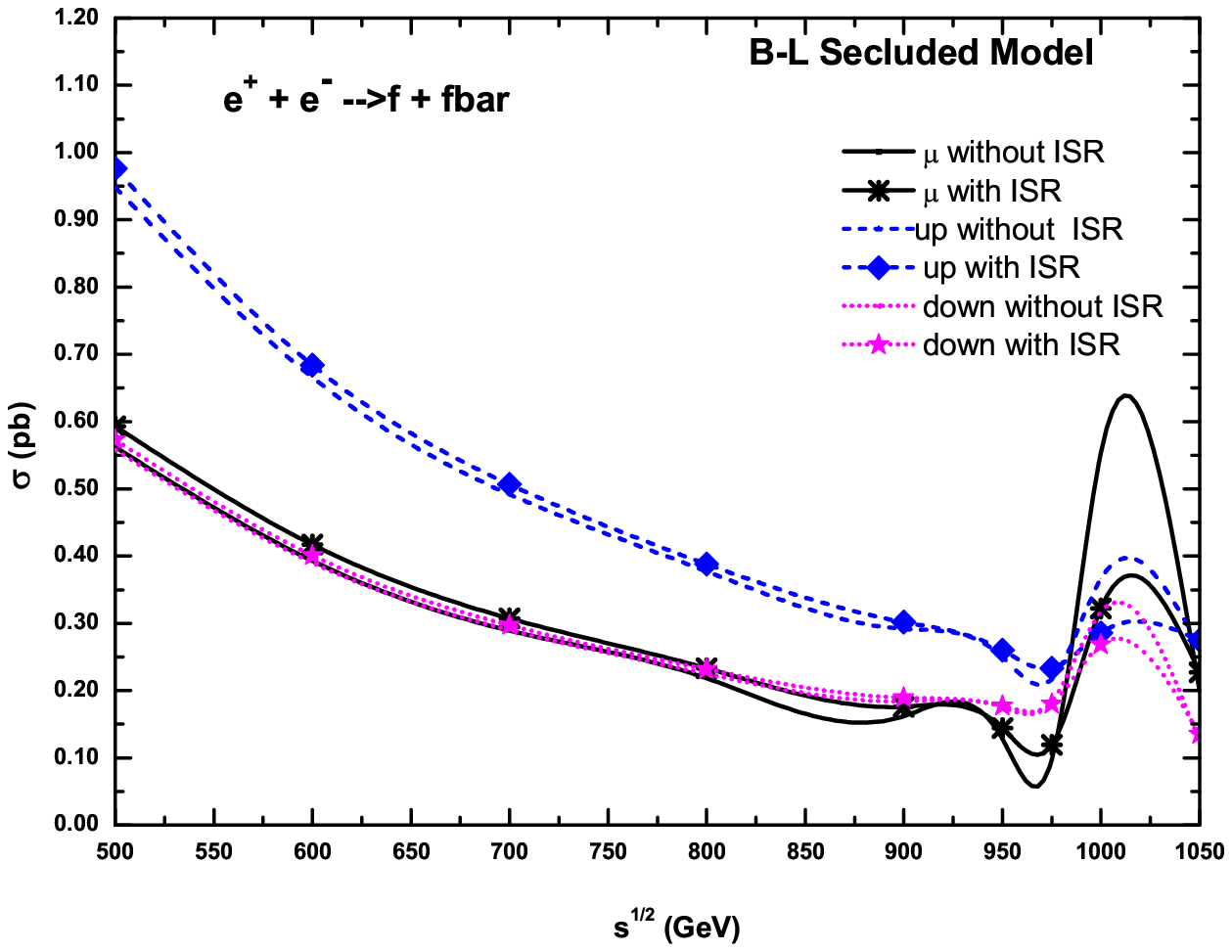}
\caption{\label{figb}  Same as in
Fig.~\ref{figa} in the secluded model.}
\end{center}
\end{figure}

%\figura{12.38}{9.6}{crossallsecl.eps}{\label{figb} Same as in
%Fig.~\ref{figa} in the secluded model.%{\color{red}Ver observa\c{c}\~{a}o }
%}

\section{Results}
\label{sec:results}

Here we will compare the models described in
Secs.~\ref{subsec:flipped}, \ref{subsec:secluded} and \ref{subsec:e6lr} by calculating
the several asymmetries, defined below, in $e^+e^-\to f\bar{f}$  processes at the TeV
energy scale, i.e., typical energies of the ILC.
We also show the prediction of the SM  contributions to these
asymmetries. The partial decay widths for both $Z_1$ and $Z_2$ bosons
were calculated, see  Appendix~\ref{sec:widths},
and the cross sections were obtained by using CompHep 4.5.1 in which the models
have been implemented. We will use the masses for the $Z_2$ given by Eqs.~(\ref{massvector1}) and
(\ref{massvector4}) of the flipped and secluded models, respectively, and choose the value of the parameters
for both models in such a way that the masses of $Z_2$ are the same in
the models: i.~e., equal to 1 TeV.  For the case of $E_6$ and LR symmetric models the total
decay widths were obtained from those of \cite{dittmar} but with a $Z_2$ mass of 1 TeV.

Concerning the mass of the extra neutral vector boson, here denoted by $Z_2$,
there are several constraints coming from direct search at hadron colliders as the Tevatron~\cite{cdf}
and from the electroweak precision tests at LEP2, and low-energy neutral current
experiments~\cite{erler}. However, none of them have considered the flipped model. In this model
we have not used the $Z-Z^\prime$ mixing angle because it is a function of the other parameters of
the model and, for this reason, it is not a free parameter anymore. See Appendix \ref{sec:neutral1}.
According to Ref.~\cite{erler} electroweak precision tests imply a lower limit of 442 GeV on the $Z_2$ mass
while there is no limit coming from Tevatron and LEP2.

The secluded model has universal vector couplings $f_V$, hence $Z_2$ behaves as a heavy photon.
The effective $Z_2$ interactions that are added to the standard model Lagrangian are
of the sort $\Lambda^+_{VV}$, in the notation of Ref.~\cite{eichten}. For these effective interactions the strongest
constraint comes from measurements of $e^+e^-\to l^-l^+$ above the $Z$ peak at the LEP2 and they imply
\cite{carena}
\begin{equation}
M^2_2\geq\frac{g^2_z}{4\pi}(\Lambda^+_{VV})^2,
\label{hurra}
\end{equation}
where $\Lambda^+_{VV}$ is the energy scale at which new physics would appears
having vector couplings to the SM leptons.
LEP2 found $\Lambda^+_{VV}=21.7$ TeV, see Table 8.13 of~\cite{lepcoll}.
Then the constraint $M_2/g_z>6$ TeV arises~\cite{carena}.
There are also constraints from precision tests using the oblique parameters however they
include only the degrees of freedom of the SM~\cite{caccia}. However, in the present models there are
also right-handed neutrinos which contributions can enhance or diminish those parameters. Other
theoretical analysis consider models like the secluded $B-L$ with three parameters, $M_2,g_z$, and $k$, where $k$
denotes a mixing in the kinetic term~\cite{salvioni}. However, we have worked both models in a basis in which
the $k$ parameter has been already transformed away. For simplicity, in this work we use $Z_2$ masses of
1 TeV, just for illustrating the possible behavior of the models when they are compared to each other through
the study of several asymmetries. Thus, we will chose the free parameters of the models
$g^\prime,g_{_{B-L}}$ and the VEV $u$ in the flipped model, and $g_z$ and $u$ in the secluded one, in such a
way that the $Z_2$ mass is the same in both models. For the other models we have used the parameters shown
in Table~\ref{table1}. On the other hand, in the ILC with CM energy of 1 TeV,
6-20 TeV resonances may still be discovered by their virtual effects~\cite{desh1}.

\subsection{Decay Widths}
\label{subsec:dw}

The total $Z_2$ decay width is given by $\Gamma^T_2=3(\Gamma_{\nu\nu}+\Gamma_{\nu_R\nu_R}+
\Gamma_{ll}+\Gamma_{dd})+2\Gamma_{uu}+\Gamma_{tt}+\Gamma_{WW}+\sum_i\Gamma_{Z_1H_i}$.
The partial decay widths $\Gamma_{Z_1H_i}$ for scalar masses of some
hundred GeV are negligible and $\Gamma_{WW}$
is different from zero only in the flipped model. For the secluded model,
at the tree level and also at the 1-loop level, $Z\equiv Z_1$ and $Z^\prime\equiv Z_2$; thus,
$Z^\prime$ has no couplings with the $W$ bosons and $Z_1$ does not couple to right-handed neutrinos
and neither to the scalar singlet. In this work, we will also consider that, in the flipped model,
the $Z_2$ decay in two right-handed neutrinos are kinematically forbidden. Below we will calculate
several asymmetries using a set of values for the parameters in these models. First, let us say in
what conditions the several observables and pseudo-observables were calculated.

For the flipped $B-L$ model of Sec.~\ref{subsec:flipped} we use
the following inputs: $g_{_{B-L}}=0.6132$, $g^\prime=0.44$,
$u=1324.4$ GeV and $v=246$ GeV.  We have also used $\alpha=1/127.9$ and
$s^2_W=0.23122$~\cite{lhcilc}.
The total $Z_2$ decay width is $\Gamma^T_{Z_2}=18.85$ GeV, and the most important
branching ratios are: $3BR(Z_2\to\nu\bar{\nu})\approx42.92\%$, $3B(Z_2 \to l^-l^+)
\approx16.17\%$, $2BR(Z_2\to u\bar{u})\approx3.83\%$,
$3B(Z_2\to d\bar{d})\approx32.49\%$, $B(Z_2\to t\bar{t})\approx 1.51\%$, and $BR(Z_2\to
W^{+}W^{-})\approx3.07\%$. As we said before, the partial decay widths $BR(Z_2\to ZH_{1,2})$ are
negligible ( $H_1$ is the neutral scalar which is almost a doublet $m_{H_1}=115$ GeV, and $H_{2}$
is the neutral scalar which is almost singlet~\cite{blsm} and we have assumed $m_{H_2}=484.73$ GeV.)

For the secluded $B-L$ model of Sec.~\ref{subsec:secluded} we use two inputs:
$g_z=0.2$, and $u=5$ TeV  and the other
parameters are as in the flipped model. With these parameters  we obtain $M_2=1$ TeV.
The values of the neutral current coupling constants, from
Eq.~(\ref{fvfazh0}), are also shown in Table~\ref{table2}.
In this model the partial widths are
$3BR(Z_2\to\nu\bar{\nu})\approx37.51\%$, $3BR(Z_2\to l^-l^+)\approx37.51\%$,
$2BR(Z_2\to u\bar{u})\approx8.34\%$,
$3BR(Z_2\to d\bar{d})\approx12.50\%$, and
$BR(Z_2\to t\bar{t})\approx4.14\%$.
The values that we have used for the mass of $H_2$ are the same as in the previous
model.

\vskip .5cm
\begin{table}[ht]
\begin{center}
\begin{tabular}{|l||c|c||c|c|}
\hline
& \multicolumn{2}{|c||}{Flipped } &
\multicolumn{2}{|c||}{Secluded}  \\ \cline{2-5}
  & $f_V$ & $f_A$  & $f_V$ & $f_A$\\\hline
neutrinos & 0.841  &  -0.174   & 0.269   &  0  \\
leptons   & 0.498  & 0.174        & 0.269    &   0  \\
$u$-quarks & -0.051 & -0.174        &-0.089 &   0 \\
$d$-quarks & -0.395  & 0.174       & -0.089 &0 \\
\hline
\end{tabular}
\end{center}
\vskip -0.5cm
\caption{Values for the neutral coupling constants $f_{V,A}$ in the
flipped and secluded models for the value of the parameters that are shown in the text.}
\label{table2}
\end{table}

The values of $f_{V,A}$ for the inputs for the flipped and secluded models are shown in Table \ref{table2}.
In both models, as noted in the case of the secluded model in Ref.~\cite{basso}, the leptonic
branching ratios of the $Z_2$ are greater than those of the
$Z$ of the SM (which are about 10\%).

For the $Z^\prime_\chi$, $Z^\prime_\psi$, $Z^\prime_\eta$ and LR models we have the total widths (in GeV):
$ 16.07,\, 8.73,\, 9.60$, and $25.73$, respectively for a $Z^\prime$ mass of 1 TeV and the couplings
can be obtained from Table~\ref{table1} using $\beta=0$, $\beta=\pi/2$ and $\beta=\arctan(-\sqrt{5/3})$ for the $U(1)_\chi,
U(1)_\psi$ and $U(1)_\eta$ models, respectively, and $\alpha_{LR}=\sqrt{2}$ for the LR symmetric model~\cite{dittmar}.

In the $B-L$ flipped model the  $Z_2$ total decay width is larger than in the secluded one.
We consider partial decay widths with and without leading QED and QCD
corrections i.~e., for the case of QED, there is a factor $1+(3\alpha/4\pi) (Q^f)^2$ for
charged fermions, and $1+\alpha_s(M^2_Z)/\pi$ (QCD) for the light quarks,
in the final states, see Leike in Ref.~\cite{zprime}, but for the top quark we used $\alpha_s(m^2_t)$.
No electroweak corrections have been
considered, and the asymmetries were calculated in the Born approximation.
The evolution of the total decay width of the $Z_2$ with $M_2$, in both
models, is shown in Fig.~\ref{fig1}. The total $Z_2$ decay widths  with and without
QED and QCD corrections are shown in Table~\ref{table3}.

\begin{table}
%\begin{eqnarray*}
\begin{tabular}{|c||c|c|c|}
  \hline
  % after \\: \hline or \cline{col1-col2} \cline{col3-col4}
  \phantom{...}
 & \textrm{nrc} & \textrm{rc} &
 \textrm{variation}\, (\%) \\ \hline
 \textrm{Flipped (1)}  & 18.85     & 19.13& 1.46\% \\
 \hline
 \textrm{Secluded (1)} & 2.12 & 2.14 & 0.93 \%    \\ \hline
%\end{array}
\end{tabular}
\caption{Total $Z_2$ widths, in GeV, for the flipped and secluded models without (nrc) and
with (rc) radiative corrections using the inputs given in the text. }
\label{table3}
\end{table}

It is well known that in the SM the $Z$ couplings to left- and
right-fermions are different and this implies several asymmetries
that were measured by LEP and SLD with high precision~\cite{lepsld}.
The same happens in models with extra electrically neutral vector bosons.
However, the secluded $B-L$ model has a heavy vector boson which couples
to fermions only through vector couplings. As we said before, it behaves like a heavy photon,
but it also has interference with the photon and $Z$, and its effects are
visible in the asymmetries. Although we have calculated the asymmetries for all the quarks,
we only show in the figures the forward-backward asymmetry for the top quark final states.
Before showing the analysis of the asymmetries we consider the cross sections.

\subsection{Cross sections and the number of events}
\label{subsec:crosssections}
%{\color{red}Acredito que grande parte dessa se\c{c}\~{a}o pode ser retirada. Ver comentario no e-mail.}

We study the production of the extra neutral gauge bosons $Z_2$ with mass of
1 TeV in the context of the ILC, which is supposed to start working with an energy of 0.5 TeV
and a possible energy upgrade to 1 TeV is being planned. Here we show the cross sections and the number of events
only for the flipped and secluded models. This is because the flipped model, as we have considered here, has not
been studied in the literature, and the most similar model that has already been well studied is the secluded one.
We have imposed cuts: the invariant mass has to be greater than 100 GeV, and the
cosine of the angle between the incoming and the outgoing fermions obeys $-0.99 < \cos\theta < 0.99$.

\begin{table}[ht]
\begin{center}
\begin{tabular}{|l||c|c||c|c|}
\hline
& \multicolumn{2}{|c||}{Flipped} & \multicolumn{2}{|c|}{Secluded} \\ \cline{2-5}
  & without BC & with BC   & without BC  & with BC \\
\hline
leptons & 1.894 & 1.126  &0.553 	  & 0.322	      \\
$u$-quarks &0.870  & 0.612  &  0.366   &    0.286     \\
$d$-quarks & 3.950  & 2.654 &0.317  &    0.268 \\
\hline
\end{tabular}
\end{center}
\vskip -0.5cm
\caption{Total cross sections (in pb) for $e^+e^-\to f\bar{f}$ with
$f=\mu,\tau$, $u$-,$d$-like fermions in the flipped and secluded models, with and without BC, at the $Z_2$ peak.}
\label{table4}
\end{table}

The ILC combination of high energy and high luminosity per bunch gives
rise to beam-beam interaction like disruption, beamstrahlung and
coherent pair production~\cite{dvs}. We will also try to give an
estimation of the beamstrahlung effects, which are the main source of beam
related backgrounds. The energy loss due to beamstrahlung is, for the ILC, approximately
equivalent to the initial state radiation.
Thus, we have to include them in our calculations. In our simulations, we have considered
for each two beams that a bunch has rms dimensions $\sigma_y=5.7$ nm high, $\sigma_x=655$
nm wide and horizontal beam size $\sigma_z= 300$ $\mu$m, and contains $2 \times 10^{10}$ particles
according to Ref.~\cite{hitoshi}.

For all the calculations, we have used the exact neutral
current couplings given in Appendix~\ref{sec:neutral1} for the
flipped model, and Eq.~(\ref{fvfazh0}) for the secluded model. We show in Table~\ref{table2}
their respective numerical values,
for the parameters given in Sec.~\ref{subsec:dw}. We use
kinematic constraints when necessary and we have neglected the mass for all the
fermions, except for the top quark, and we have also used the unitary gauge.

%%%
\begin{figure}[ht]
\begin{center}
\includegraphics[height=.5\textheight]{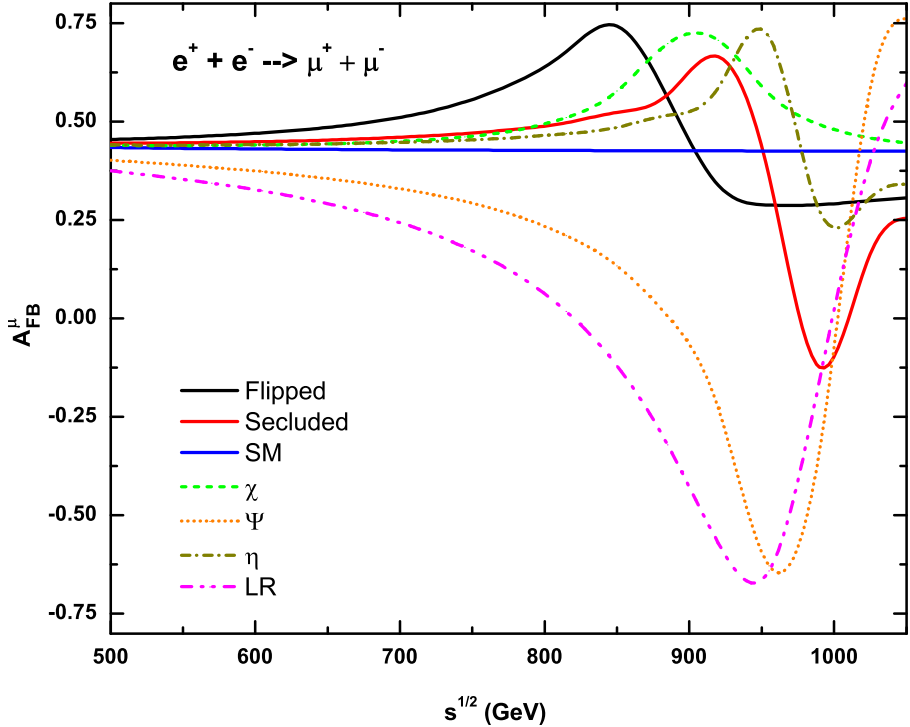}
\caption{\label{fbmuon} Forward-backward %fig6a
asymmetry for the $e^+e^-\to \mu^-\mu^+$ in the flipped,
secluded, $U(1)_{\chi,\psi,\eta}$ and LR models. We also show the SM contribution.}
\end{center}
\end{figure}

The expected integrated luminosity for the ILC in the first four years of
running is 500 $\textrm{fb}^{-1}$. In our simulations we consider an integrated luminosity
of 100 $\textrm{fb}^{-1}$ in a year~\cite{fnal}. The total cross sections (pb) given in
Table \ref{table4} are shown with and without beamstrahlung corrections (BC).
The flipped model if compared to the secluded one has a higher number of events, by a factor 10.
We can realize that the beamstrahlung effect decreases the cross section in the region around the $Z_2$ peak,
as we can see from Figs.~\ref{figa} and \ref{figb}.  We can also see from these figures that the secluded model (mainly
without beamstrahlung corrections) has cross sections near the $Z_2$ peak, with leptons in the final state,
that are larger than the cross section for quarks. However, these features depend on the chosen parameter
values.

\begin{figure}[ht]
\begin{center}
\includegraphics[height=.5\textheight]{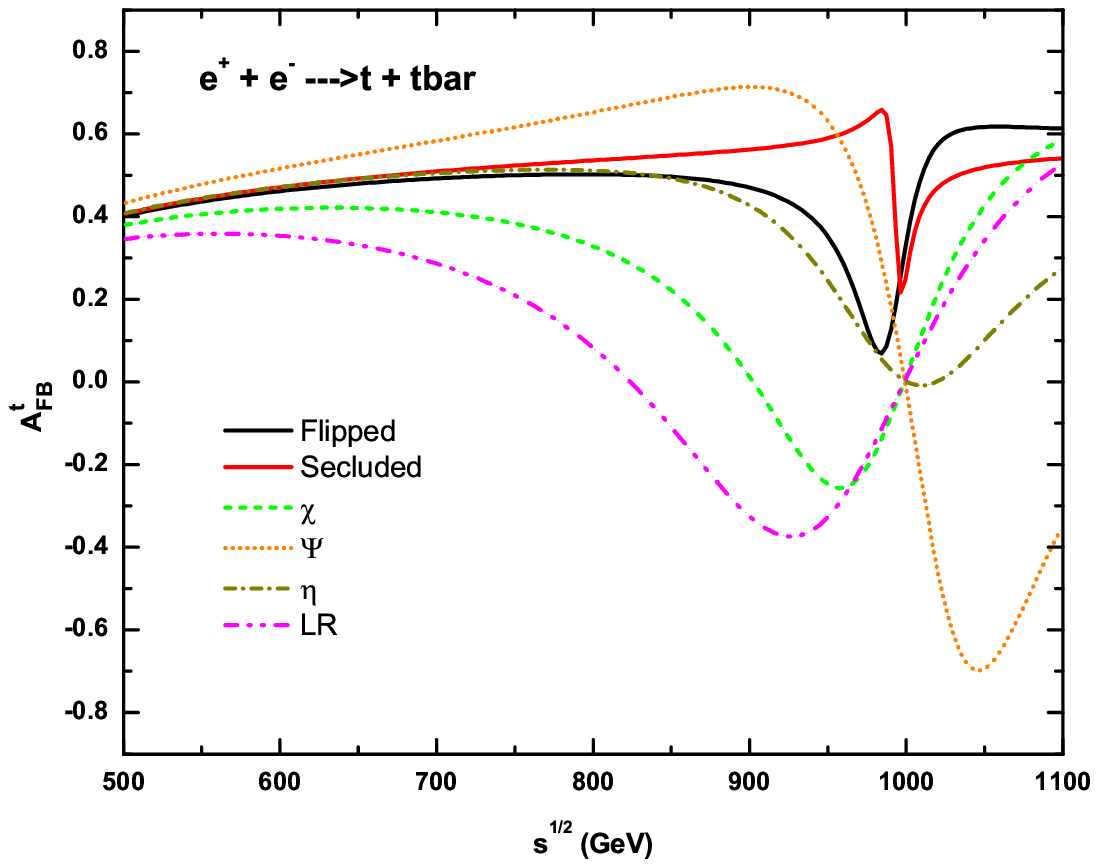}
\caption{\label{fbtop}Forward-backward %fig6
asymmetry for the $e^+e^-\to t\bar{t}$ process in the flipped,
secluded, $U(1)_{\chi,\psi,\eta}$ and LR models.}
\end{center}
\end{figure}

%%%
\begin{figure}[ht]
\begin{center}
\includegraphics[height=.5\textheight]{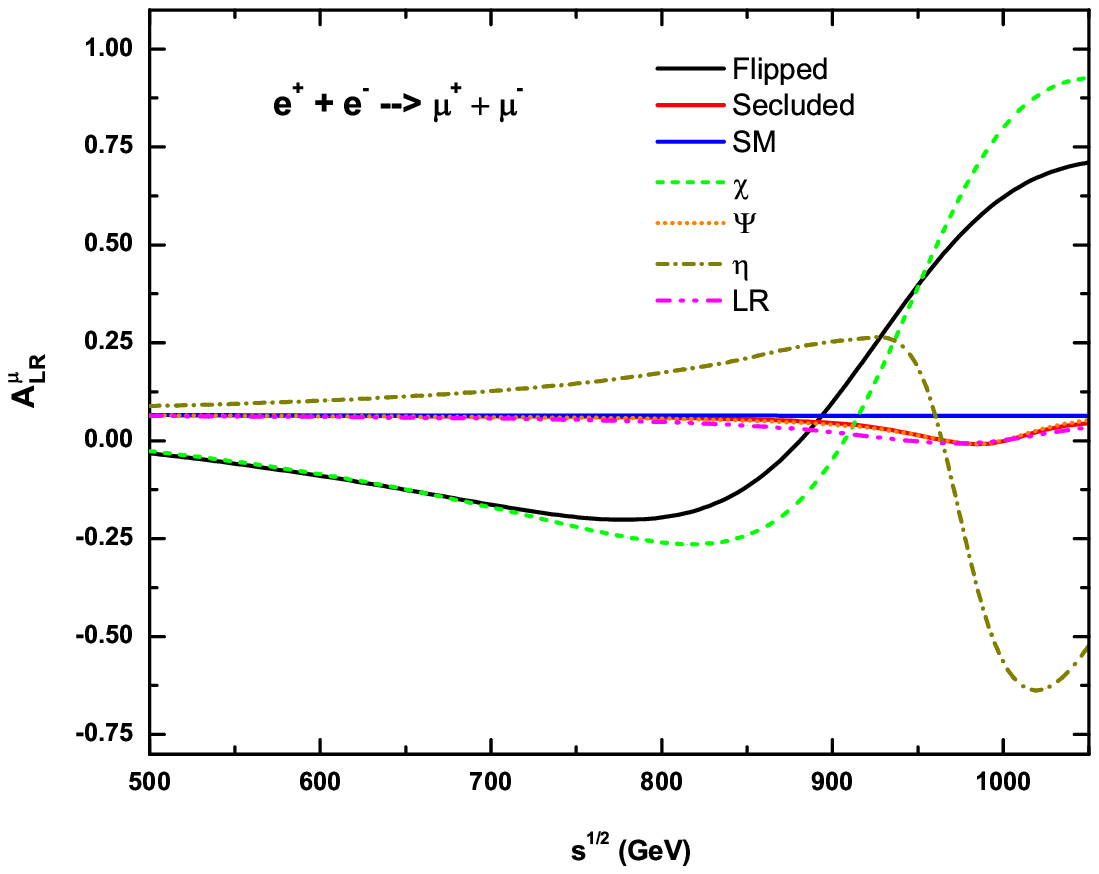}
\caption{\label{lrmuon}Left-right %fig6b
asymmetry for the $e^+e^-\to \mu^-\mu^+$  in the flipped,
secluded, $U(1)_{\chi,\psi,\eta}$ and LR models.
We also show the SM contribution.}
\end{center}
\end{figure}

\subsection{Asymmetries}
\label{subsec:assi}

In this subsection we show the results of several asymmetries for the models
considered in the previous section. The considered asymmetries are:
the forward-backward, left-right, polarization, mixed left-right-forward-backward, and mixed
forward-backward-polarization asymmetries.
The cross sections for elementary particles contain  terms coming from interactions
among all bosons, $\gamma,Z,Z^\prime$. Thus, we have three terms of resonance
and three terms of interference of $\gamma-Z$,$\gamma-Z^\prime$ and $Z-Z^\prime$. The behavior of
the asymmetries depends on the relative magnitude of these terms, which in turn depend on the energy scale.

The forward-backward asymmetry for the fermions $i\not= e$ in the final states, is defined as
\begin{equation}
A^{i}_{FB}=\frac{\int_0^1dz\,(d\sigma^i/dz)-\int_{-1}^0dz\,
(d\sigma^i/dz))}{\int_{-1}^1dz\,(d\sigma^i/dz)}\equiv \frac{\sigma^{i}_F-\sigma^{i}_B}{\sigma^i},
\label{alr}
\end{equation}
where $\sigma^i$ is the cross section of the process $e^+e^-\to f_i\bar{f_i}$ and $z=\cos\theta$, where
$\theta$ is the angle between the arriving electron and the final state fermion. The latter
expression defines $\sigma^i_{F(B)}$.

The left-right asymmetry is defined as
\begin{equation}
A^{i}_{LR}=\frac{\sigma(e^+e^-_L\to f_i\bar{f}_i)-
\sigma(e^+e^-_R\to f_i\bar{f}_i)}{\sigma(e^+e^-\to
f_i\bar{f}_i)}\equiv \frac{\sigma^i_L-\sigma^i_R}{\sigma^i},
\label{alr2}
\end{equation}
where $e^-_{L(R)}$ denote the left-(right-) handed
longitudinally polarized electrons. We are neglecting the
lepton masses and assuming 100\% of polarization. The latter expressions define $\sigma^i_ {L(R)}$.

The polarization asymmetry is defined as
\begin{equation}
A^{i}_{pol}=\frac{\sigma^i_R-\sigma^i_L}{\sigma^i},
%}\equiv \frac{(\sigma^i_{LL}+\sigma^i_{RL})-(\sigma^i_{LR}+\sigma^i_{RR})}
%{\sigma^i_{LL}+\sigma^i_{RL}+\sigma^i_{LR}+\sigma^i_{RR}},
\label{assipol1}
\end{equation}
where now the subscripts refer to the helicities outgoing fermion.
It was already measured at LEP~\cite{etzion}, by studying the polarization of the
$\tau^+\tau^-$ in some decay
channels like: $\tau^-\to \pi^-\nu_\tau,\rho^-\nu_ \tau$, and
$\tau^-\to\,l^-\bar{\nu}_l\nu_\tau$. At ILC this asymmetry can be  measured also for
$t\bar{t}$ final states in the channels $t\to bW^+\to bl^+\nu_l,bu\bar{d},bc\bar{s}$~\cite{leike2,arai}.

%%%
\begin{figure}[ht]
\begin{center}
\includegraphics[height=.5\textheight]{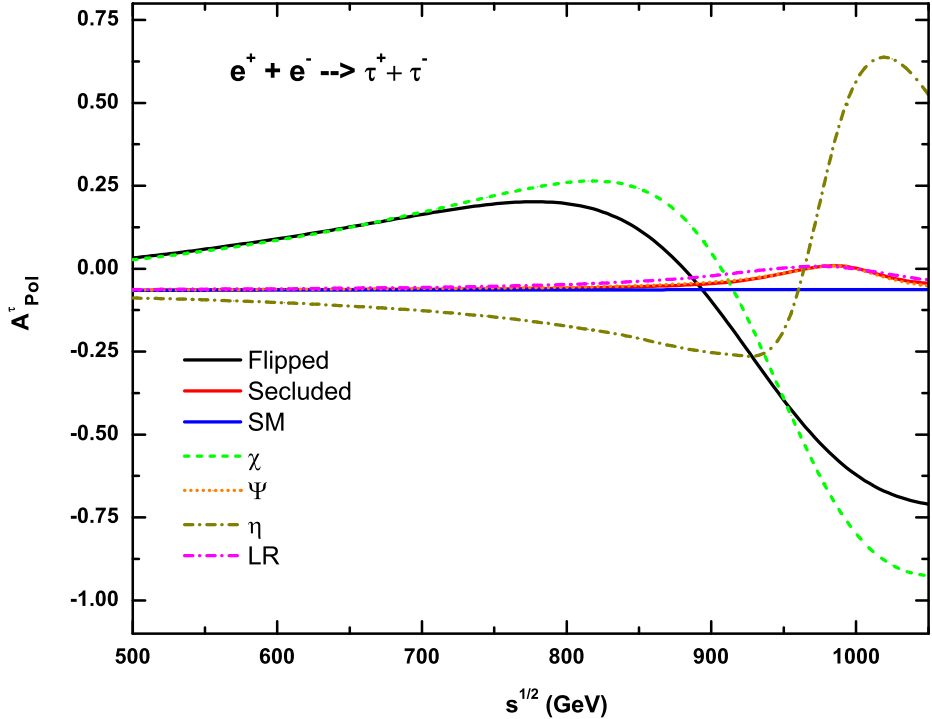}
\caption{\label{poltau} Polarization %fig8a
asymmetry for the $e^+e^-\to \tau^-\tau^+$  in the flipped,
secluded, $U(1)_{\chi,\psi,\eta}$ and LR models. We also show the SM contribution.}
\end{center}
\end{figure}

The $LRFB$ asymmetry is defined as
\begin{equation}
A^i_{LRFB}
=\frac{(\sigma^{i}_F-\sigma^{i}_B)_L-(\sigma^{i}_F-\sigma^{i}_B)_R}{\sigma^i},
\label{mixedassi1}
\end{equation}
this combined asymmetry can lead to a statistical precision that is equivalent to the unpolarized
forward-backward asymmetry, depending on the level of polarization achieved.
%%%%
\begin{figure}[ht]
\begin{center}
\includegraphics[height=.5\textheight]{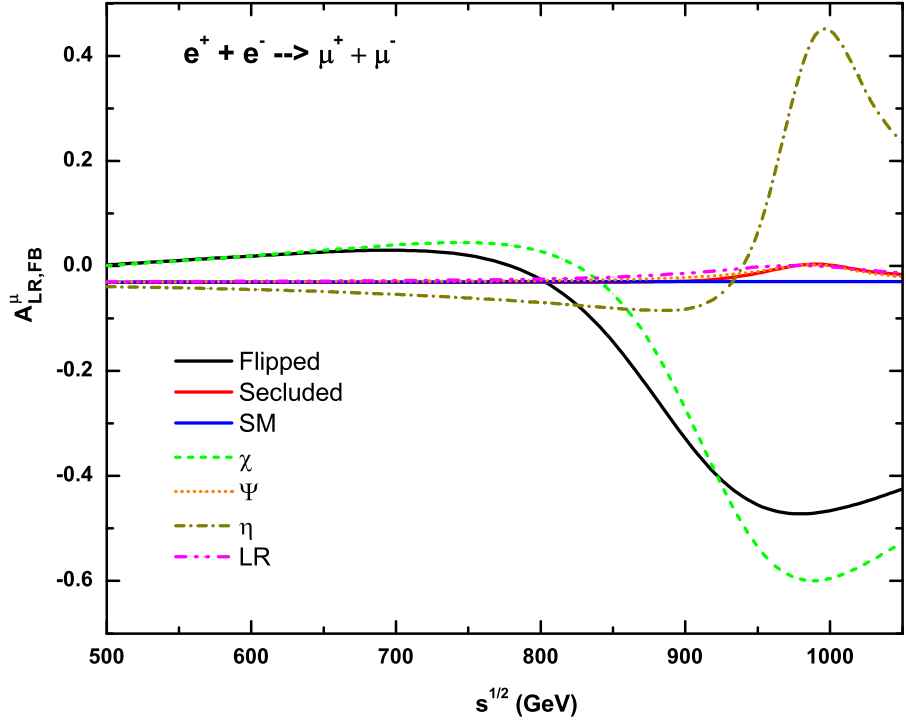}
\caption{\label{fblr} Combined $LRFB$ %fig11a
asymmetry for the $e^+e^-\to \mu^-\mu^+$  in the flipped,
secluded, $U(1)_{\chi,\psi,\eta}$ and LR models.
We also show the SM contribution.}
\end{center}
\end{figure}

There is another mixed asymmetry, the
polarized forward-backward double asymmetry,  $A^i_{FB}(pol)$, which is
defined as follows~\cite{lepsld}:
\begin{equation}
 A^{i}_{FB}(pol)=\frac{(\sigma^{i}_R- \sigma^{i}_L)_F-(\sigma^{i}_R-\sigma^{i}_L)_B}
{\sigma^i},
\label{newa}
\end{equation}
where $(\sigma^{i}_L)_F$ denotes the cross section in the forward direction, with the fermion in the final state, $f_i$,
left-handed polarized for the $e^+e^-\to f_i\bar{f}_i$ process, etc.

\begin{figure}
\begin{center}
\includegraphics[height=.5\textheight]{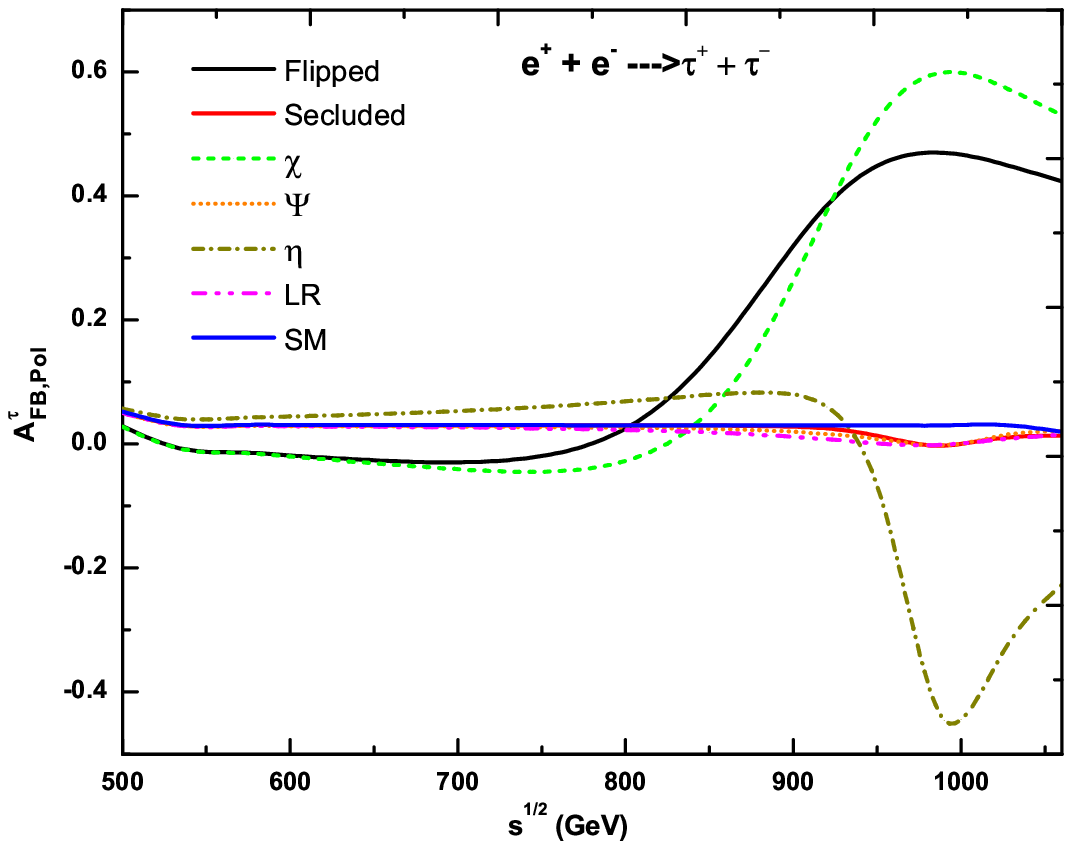}
\caption{\label{fbpol}Combined $FB(Pol)$ %fig12
asymmetry for the $e^+e^-\to \tau^-\tau^+$  in the flipped,
secluded, $U(1)_{\chi,\psi,\eta}$ and LR models.
We also show the SM contribution.}
\end{center}
\end{figure}

\subsection{Discussion}
\label{subsec:discu}

Since we are neglecting the lepton masses, the asymmetries for $\mu^+\mu^-$ and $\tau^+\tau^-$
are the same, but we have labeled them differently to point out that the forward-backward and left-right
asymmetries can be measured by detecting muons, and  asymmetries involving polarization can only be measured
by detecting the $\tau$ decay product polarizations. On the other hand, for observables involving the top quark
we take its mass into account. The asymmetries involving polarization in the latter case
will be show elsewhere.

\begin{table}
%\begin{eqnarray*}
\begin{tabular}{|c||c|c|c|c|c|c|c|}
  \hline
  % after \\: \hline or \cline{col1-col2} \cline{col3-col4}
  \phantom{...} &  Flipped& Secluded & $U(1)_\chi$  & $U(1)_\psi$ & $U(1)_\eta$ & LR & SM\\ \hline
  $A^{(0,\mu)}_{FB}$ & 0.293& $2.65\times10^{-4}$ &0.479 &$4.8\times10^{-3}$ &0.272 & $12.6\times10^{-3}$ & 0.425\\
 $A^0_{LR}$  & 0.621   & $3.16\times10^{-4}$&-0.798 &$0.548\times10^{-3}$  & -0.593&$1.4\times10^{-3}$ &0.064\\
 $A^0(pol)$ & -0.621 &$-0.316\times10^{-4}$ & -0.798 &$-0.548\times10^{-3}$&0.593&$-1.4\times10^{-3}$  & -0.064 \\
 $A^0_{LRFB}$ &-0.466 &$-0.148\times10^{-4}$&-0.598 &$-0.259\times10^{-3}$&0.445&$-0.7\times10^{-3}$ & -0.030\\
 $A^0_{FB}(pol)$ & 0.466&$0.148\times10^{-4}$ & 0.598&$0.259\times10^{-3}$  & -0.445& $0.7\times10^{-3}$ &0.030
  \\
 \hline
%\end{array}
\end{tabular}
\caption{Values of the several asymmetries at the $Z_2$ peak, $\sqrt{s}=1$ TeV.  }
\label{table5}
\end{table}
The asymmetries that we consider in this work can be used to discriminate some of
the models, but in some cases some ambiguities remain. Their behavior depend on the values
of the neutral couplings $f_{V,A}$ in each model. The $E_{6}$ models, $U(1)_\chi$, $U(1)_\psi$ and
$U(1)_\eta$, have $f^u_V=0$, $U(1)_\psi$ has also $f^l_V=0$, and the left-right symmetric model has
$f_{V}^l=0$ if $\alpha_{LR}=\sqrt2$. On the other hand, the $B-L$ secluded model has $f_A=0$ for all
fermions. In Figs.~\ref{fbmuon}-\ref{fbpol} we show the asymmetries in the context of
the models considered above and in Table~\ref{table5} we show their values at the
$Z_2$ peak.

In Fig.~\ref{fbmuon} we show the
forward-backward asymmetry for muons in the final state. The LR and the
$U(1)_\psi$ models can be easily discriminated from the other ones: $U(1)_\chi,U(1)_\eta$,
secluded, and flipped models, and we also note that both of them present a similar
behavior. The same asymmetry with the top quark in the final state, see Fig.~\ref{fbtop},
allows to distinguish the LR from the $U(1)_\psi$ model and, both of them
from the secluded, flipped and the $U(1)_\eta$ models.

The left-right asymmetry is shown in Figs.~\ref{lrmuon}.  We see from that figure that the flipped and the
$U(1)_\chi$ models have similar behavior and, that the secluded, $U(1)_\psi$, and LR models are difficult
to discriminate even from the SM contribution. The $U(1)_\eta$ has a different behavior. For $\sqrt{s}>800$ GeV
it can be distinguished from the other models.

For the polarization asymmetry, see Fig.~\ref{poltau}, we have only considered massless fermions and, because of the
universality of the coupling constants, this asymmetry is just the left-right asymmetry with a minus
sign when the final states are leptons.

The $LRFB$ asymmetry is shown in Fig.~\ref{fblr}.  The qualitative behavior of this asymmetry is similar
to the behavior of  polarization asymmetry in Fig.~\ref{poltau} however, the respective
values for the $LRFB$ asymmetry are smaller than those of the polarization
asymmetry in Fig.~\ref{poltau}. On the peak the $LRFB$ asymmetry
is equal to $(3/4)A_{Pol}$.

Next, we analyze the forward-backward polarized asymmetry that is shown in Fig.~\ref{fbpol}. From it we
see that the flipped and the $U(1)_\chi$ models can be discriminated from the $U(1)_\eta$ but the other models
present signal almost equal of that of the SM. This asymmetry gives the same information of
left-right asymmetry and considering absolute values we can also say that it
also gives the same information of polarization and $LRFB$ mixed asymmetry.

Finally, in Table \ref{table5} we show the values of the asymmetries at the $Z_2$ peak.  If the measured values are
compatible with zero, they will be in favor of the secluded, $U(1)_\psi$ and LR models. For any values different from zero,
the models flipped, the $U(1)_\chi$ or the $U(1)_\eta$ will be favored.

Some of the asymmetries considered above were also studied as functions of $M_2$ in Ref.~\cite{godfrey2}.

\section{Conclusions}
\label{sec:con}

If a new neutral vector boson exists with a mass of the order of TeVs it could be
discovered at  the LHC. If this happens,
the next task will be to measure its parameters, i.~e., its mass, spin, couplings
to fermions, etc. Some of these parameters may be obtained by measuring asymmetries
like those considered in this paper, on or/and off peak. Different models will have
different behavior for these asymmetries, at least in some range of $\sqrt{s}$.
In this paper we have shown that this, in fact, happens with the models we have considered above.
Our main results are the calculation of cross sections, decay widths
and asymmetries in the Born approximation for the flipped and secluded models. The asymmetries,
have also been calculated in $E_6$ inspired and left-right symmetric models.
The flipped and secluded considered here have at least an extra heavy scalar and three
right-handed heavy neutrinos, and the LR model has besides the right-handed neutrinos, several
scalar fields. Hence, in these models these degrees of freedom have to be taken into
account when calculating the respective radiative corrections.
However, the extra parameters in the models are not known at
present and thus it is not worth carrying out those
calculations. Notwithstanding, some corrections are more
general like those of QED and QCD, which in these models are the same of the SM, and we have
taken them into account. On the other hand, it is always important to
know what is the value of a given observable in a particular model, at least at the tree level.
Only then can we appreciate the importance of the radiative corrections.

As we said before, the flipped model as we have considered here has not been studied in the literature; thus,
we have given more details and have compared it with a model which is very similar with it: the secluded model.
Both models have similar quantum numbers and degrees of freedom. In particular we note that
the secluded model near the $Z_2$ peak has cross sections $\sigma(e^-e^+\to f\bar{f})$ for charged leptons
larger than the cross section for quarks. The $Z_2$ decay widths
are very different in each model and are larger in the flipped one.
For asymmetries it is not necessary to be at the $Z_2$ peak, here at 1000 GeV. The
presence of a heavy $Z_2$ may be detected well before the peak, allowing also the discrimination
of each model.

\acknowledgements E. C. F. S. Fortes was supported by FAPESP
under contract No. 2007/59398-2; J. C. Montero and V. Pleitez were partially
supported by CNPq under Contract Nos. 302102/2008-6
(JCM) and 302102/2008-6 (VP). We also would like to thank an anonymous referee for valuable comments
which improved this paper from the submitted version.

\appendix

\section{The fermion-vector boson interactions in the flipped
model}
\label{sec:neutral1}

We will parameterize the neutral current couplings of a fermion $\psi_i$ as
in Eq.~(\ref{nc}) where the exact expressions for
$Z_1$ and $Z_2$ are used in defining $g^i_{V,A}$ and
$f^i_{V,A}$. We also show them in the approximation $\bar{v}\ll1$ in the text,
see Eqs.~(\ref{nusnosso})-(\ref{dnosso}).

The couplings of the neutrinos are:
\begin{eqnarray}
g^\nu_V&=& \frac{1}{2}\sqrt{\frac{t^{\prime\,2}+t^2_{_{B-L}} }{
2B(C-D\sqrt{X})}}\, [E+(1-t^{\prime\,2})\sqrt{X}], \nonumber \\
g^\nu_A&=&-\frac{1}{2}\sqrt{\frac{t^{\prime\,2}+t^2_{_{B-L}}
}{2 B(C-D\sqrt{X})}}\, [D -(1+t^{\prime\,2})\sqrt{X}],
\nonumber \\
f^\nu_V&=&\frac{1}{2}\sqrt{\frac{t^{\prime\,2}+t^2_{_{B-L}}}{2B(C+D\sqrt{X})}}\,
[E-(1-t^{\prime\,2})\sqrt{X} ], \nonumber \\
f^\nu_A&=&-\frac{1}{2}\sqrt{\frac{t^{\prime\,2}+t^2_{_{B-L}}}{2B(C+D\sqrt{X})}}\,
[D+(1+t^{\prime\,2})\sqrt{X}].
\label{nusnosso2}
\end{eqnarray}

For the case of the charged leptons:
\begin{eqnarray}
g^l_V&=& -\frac{1}{2}\sqrt{
\frac{t^{\prime\,2}+t^2_{_{B-L}}}{2B(C-D\sqrt{X})}}\,
[E-32t^{\prime\,2}t^2_{_{B-L}}+
(1-t^{\prime\,2})\sqrt{X}],\quad
g^l_A=-g^\nu_A, \nonumber \\
f^l_V&=&-\frac{1}{2}\sqrt{\frac{t^{\prime\,2}+t^2_{_{B-L}}}{2B(C+D\sqrt{X})}}\,
[E-32t^{\prime\,2}t^2_{_{B-L}}-(1-t^{\prime\,2})\sqrt{X}], \quad
f^l_A=-f^\nu_A.
\label{lnosso2}
\end{eqnarray}

In the quark sector we obtain, for the $u$-like quarks:
\begin{eqnarray}
g^u_V&=&\frac{1}{6}
\sqrt{\frac{t^{\prime\,2}+t^2_{_{B-L}}}{2B(C-D\sqrt{X})}}\,[F+3(1-t^{\prime\,2})\sqrt{X}] ,
\quad g^u_A= g^\nu_A,
\nonumber \\ f^u_V&=&
\frac{1}{6}\sqrt{\frac{t^{\prime\,2}+t^2_{_{B-L}}}{2B(C+D\sqrt{X})}}\,[F
-3(1-t^{\prime\,2})\sqrt{X}],\quad
f^u_A=f^\nu_A.
\label{unosso2}
\end{eqnarray}

and, for the $d$-like quarks:
\begin{eqnarray}
g^d_V&=&-\frac{1}{6}\sqrt{\frac{t^{\prime\,2}+t^2_{_{B-L}}}{2B(C-D\sqrt{X})}}\,
[F+32t^{\prime\,2}t^2_{_{B-L}}+3(1-t^{\prime\,2})\sqrt{X}],\quad
 g^d_A=-g^\nu_A,
%\frac{1}{2}\sqrt{\frac{t^{\prime\,2}+t^2_{_{B-L}}}{2B(C-D\sqrt{X})}}\,[D-(1+t^{\prime\,2})\sqrt{X}],
\nonumber \\ f^d_V&=&
-\frac{1}{6}\sqrt{\frac{t^{\prime\,2}+t^2_{_{B-L}}}{2B(C+D\sqrt{X})}}\,[
F+ 32t^{\prime\,2}t^2_{_{B-L}} -3(1-t^{\prime\,2})\sqrt{X}],\quad
f^d_A=-f^\nu_A.
\label{dnosso2}
\end{eqnarray}
Where we have defined:
\begin{eqnarray}
&&C=(1+t^{\prime\,2})[16(t^{\prime\,2}+t^2_{_{B-L}})^2+8(t^{\prime\,4}-B)\bar{v}^2+
(1+t^{\prime\,2})^2\bar{v}^4],\nonumber
\\&&
D=4(t^{\prime\,4}-B)+(1+t^{\prime\,2})^2\bar{v}^2,\nonumber\\&&
E=4(t^{\prime\,4}+B)
+8t^{\prime\,2}t^2_{_{B-L}}-(1-t^{\prime\,4})\bar{v}^2,\nonumber \\&&
F=12(t^{\prime\,4}+B)
-40t^{\prime\,2}t^2_{_{B-L}}
-3(1-t^{\prime\,4})\bar{v}^2,\nonumber \\&&
X=A^2-16B\bar{v}^2. \label{notacao1}
\end{eqnarray}
and $A,B$ are defined in Eq.~(\ref{ab}).

The couplings of the charged fermions to the photon are as
usual. Notice also that, in the Eq.~(\ref{nc}) all fermions
$\psi_i$ are still symmetry eigenstates, therefore the neutral
currents coupled to $Z_1$ and $Z_2$ are flavor conserving.

%%%%%%%%%%%%%%%%%%%%%%%%%%%%%%%%%%%%%%%%%%%%%%%%%%%%%%%%%%%%%%%%%%%%5

\section{Partial $Z_{1,2}$ decay widths}
\label{sec:widths}

In both models the partial widths of the neutral vector bosons $Z_1$ and $Z_2$, respectively, are
given by:
\begin{equation}
\Gamma_{Z_1\to f_i\bar{f}_i}=\frac{N_{c}G_{F}M^2_{Z_1}}{6\pi\sqrt{2}}[(g_{V}^{i})^{2}+
(g_{A}^{i})^{2}]\sqrt{1-4\frac{m_{i}^{2}}{M_{Z_{1}}^{2}}}\left[1+2\frac{m_{i}^{2}}{M_{Z_{1}}^{2}}
\frac{[(g_{V}^{i})^{2}-2(g_{A}^{i})^{2}]}{[(g_{V}^{i})^{2}+(g_{A}^{i})^{2}]}\right]M_{Z_1},
\label{b1}
\end{equation}
and
\begin{equation}
\Gamma_{Z_2\to f_i\bar{f}_i}=\frac{N_{c}G_{F}M_{Z_{1}}^{2}}{6\pi\sqrt{2}}
[(f_{V}^{i})^{2}+(f_{A}^{i})^{2}]\sqrt{1-4\frac{m_{i}^{2}}{M_{Z_{2}}^{^2}}}
\left[1+2\frac{m_{i}^{2}}{M_{Z_{2}}^{2}}
\frac{[(f_{V}^{i})^{2}-2(f_{A}^{i})^{2}]}{[(f_{V}^{i})^{2}+(f_{A}^{i})^{2}]}\right]M_{Z_{2}},
\label{b2}
\end{equation}
where for the sake of clarity we have used here $M_{Z_1}\equiv M_1$ and $M_{Z_2}\equiv M_2$. Note that
in the flipped model the couplings of the $Z_2$ boson to the SM fermions, and the $M_{Z_2}$ mass as well,
depend on $g^\prime,g_{_{B-L}},g$ and $\bar{v}$. Only when $\bar{v}\ll1$ all of them depend only on the
gauge coupling constants. In the secluded model $f_A=0$ and we have
\begin{equation}
\Gamma^{sec}_{Z_2\to f_i\bar{f}_i}=\frac{N_{c}G_{F}M_{Z_{1}}^{2}}{6\pi\sqrt{2}}
(f_{V}^{i})^{2}\,\sqrt{1-4\frac{m_{i}^{2}}{M_{Z_{2}}^{^2}}}
\left[1+2\frac{m_{i}^{2}}{M_{Z_{2}}^{2}}\right]M_{Z_{2}}.
\label{b3}
\end{equation}
Recall that only for the $t$ quark, the fermion mass was taken into account in (\ref{b1})-(\ref{b3}).

Finally, in the flipped model the $Z_{1,2}(p)W^+(k_+)W^-(k_-)$ vertices are given by
\begin{equation}
igF_{1,2}[g_{\alpha\beta}(k_+-k_-)_\lambda+g_{\alpha\lambda}(p-k_+)_\beta+g_{\beta\lambda}(k_--p)_\alpha],
\label{z1z2ww}
\end{equation}
with all momenta incoming and where we have defined
\begin{equation}
F_1= \frac{A-2(1+t^{\prime\,2})\bar{v}^2+\sqrt{X}}{\sqrt{2Y_+}},\quad
F_2=\frac{A-2(1+t^{\prime\,2})\bar{v}^2-\sqrt{X}}{\sqrt{2Y_-}},
\label{f1fa}
\end{equation}
with $X$ defined in Eq.~(\ref{notacao1}) and we have defined
\begin{eqnarray}
Y_\pm&=&(1+t^{\prime\,2})[16(t^{\prime\,2}+t^2_{_{B-L}})^2-
8(B-t^{\prime\,4})\bar{v}^2+(1+t^{\prime\,2})^2\bar{v}^4 ]\nonumber \\&\pm&
\sqrt{X}[4(B-t^{\prime\,4})-(1+t^{\prime\,2})^2\bar{v}^2].
\label{yes}
\end{eqnarray}
where $A,B$ were defined in Eq.~(\ref{ab}) and $X$ in (\ref{notacao1}). At the tree level, in the secluded model,
the vertex $Z_1W^+W^-$ is the same as those of $ZW^+W^-$ in the SM and the vertex $Z_2W^+W^-$
does not exist. Notice that when $\bar{v}\to0$ there is no mixing between $Z$ and $Z^\prime$, thus $Z_2$ does
not decay to $W^+W^-$ at tree level and $F_2=0$. In the $E_6$ inspired models there is also no
$Z_2W^+W^-$ interaction at the tree level.

Finally, the Higgs vertices are the same as in Ref.~\cite{emam}.

%%%%%%%%%%%%%%%%%%%%%%%%%%%%%%%%%%%%%%%%%%%%%%%%%%%%%%%%%%%%%%%%

\end{document}